\documentclass[conference]{IEEEtran}

\pdfoutput=1

\usepackage{ifpdf}

\ifpdf
	\usepackage[pdftex]{graphicx}
	\graphicspath{{fig/}}
\else
	\usepackage[dvips]{graphicx}
	\graphicspath{{fig/}}
\fi

\usepackage{graphicx}
\usepackage{amsmath}
\usepackage{amssymb}
\usepackage{amsthm}
\usepackage[vlined, ruled]{algorithm2e}
	\SetKwProg{myAlgo}{Algorithm}{}{}
	\SetKwProg{myFn}{Function}{}{}
	\SetKwRepeat{Do}{do}{while}	% The above now allows you to use \Do{<end condition>}{<stuff inside the do-while loop>}

	\DontPrintSemicolon
	
	\SetCommentSty{mycommentfont}

\usepackage{multirow}
\usepackage{xcolor}
\usepackage{epstopdf}
\usepackage{verbatim}
\usepackage[multiple]{footmisc} % Multiple footnotes at one point

\usepackage[noadjust]{cite}
\newcommand{\suchthat}{\mathrel{\ooalign{$\ni$\cr\kern-1pt$-$\kern-6.5pt$-$}}}

\newcommand{\fd}[2]{\frac{d #1}{d #2}}
\newcommand{\pd}[2]{\frac{\partial #1}{\partial #2}}

\DeclareMathOperator*{\argmax}{arg\,max}

\begin{document}

\title{Joint Spectrum Reuse and Power Control for Multi-Sharing Device-to-Device Communication}


\author{
	\IEEEauthorblockN{Kuo-Yi Chen, Jung-Chun Kao, Si-An Ciou, and Shih-Han Lin}%
	%\thanks{Si-An Ciou, Jung-Chun Kao (corresponding author), Chung Yi Lee, and Kuo-Yi Chen are with the Department of Computer Science, National Tsing Hua University, Hsinchu, Taiwan. Emails: \{nsn90426, jungchuk, macaca5567, t810613t\}@gmail.com.}%
	%\thanks{Jung-Chun Kao is the corresponding author. This research was supported in part by the Ministry of Science and Technology, Taiwan, under grant no. MOST 103-2221-E-007-041-MY3.}%
	\IEEEauthorblockA{Department of Computer Science, National Tsing Hua University, Hsinchu, Taiwan\\
	Emails: \{t810613t, jungchuk, nsn90426, donyundo7642\}@gmail.com}%
}

%\author{Si-An Ciou, Jung-Chun Kao, Chung Yi Lee, and Kuo-Yi Chen%
	%\thanks{Si-An Ciou, Jung-Chun Kao (corresponding author), Chung Yi Lee, and Kuo-Yi Chen are with the Department of Computer Science, National Tsing Hua University, Hsinchu, Taiwan. Emails: \{nsn90426, jungchuk, macaca5567, t810613t\}@gmail.com.}%
	%\thanks{second author's affiliation}%
	%\thanks{third author's affiliation}%
%}

\maketitle

\begin{abstract}
	Compared to current mobile networks, next-generation mobile networks are expected to support higher numbers of simultaneously connected devices and to achieve higher system spectrum efficiency and lower power consumption. To achieve these goals, we study the multi-sharing device-to-device (D2D) communication, which allows any cellular user equipment to share its radio resource with multiple D2D devices. We jointly consider resource block reuse and power control and then develop the MISS algorithm. Simulation results show that MISS performs very well in terms of transmission power consumption, system throughput, and the number of permitted D2D devices.
\end{abstract}


\section{Introduction}

	Fifth-generation (5G) mobile networks, compared to 4G mobile networks, are expected to support higher numbers of simultaneously connected devices and achieve higher system spectrum efficiency. Meanwhile network energy efficiency shall be improved significantly. As forecast in \cite{key_parameters_for_5G}, 5G mobile networks will have 10X connection density, 2/3/5X spectrum efficiency, and 100X energy efficiency.

	To achieve these goals, \emph{multi-sharing} device-to-device (D2D) communication is a promising component \cite{Kao2015}. Unlike the \emph{single-sharing} counterpart which restricts each cellular user equipment (CUE) to share its resource block (RB) with up to one D2D user equipment (DUE), multi-sharing D2D communication allows multiple DUEs to reuse radio resource allocated to each CUE. By dynamic RB reuse within cells, multi-sharing D2D communication can increase the frequency reuse factor far greater than one.

	D2D communication \cite{Doppler2009, Yu2011} in general can reuse the uplink or downlink RBs of CUEs, but the research focus is more on \emph{uplink} resource sharing. This favoring of uplink resource sharing is mainly because asymmetric Internet traffic makes the uplink spectrum often underutilized. For single-sharing D2D communication, Han et al. \cite{Han2012} in 2012 aimed to maximize the number of permitted DUE pairs and minimize total interference under prerequisite of maximal permitted DUE pairs, with the assumption that transmission power values on all devices are predetermined and given. This max-DUE-min-interference problem in nature is the assignment problem and the proposed algorithm is essentially the Hungarian method well known for obtaining the optimal solution to the assignment problem.

	Feng et al. \cite{ORA} also studied single-sharing D2D communication but aimed to maximize the system throughput instead. Unlike \cite{Han2012} which assumes transmission power values are given, Feng et al. considered both RB reuse and power control. The Optimal Resource Allocation (ORA) algorithm they developed consists of admission control, transmission power allocation, and DUEs-to-CUEs maximum weight matching. Wang et al. \cite{Wang2013} also studied single-sharing D2D communication with joint consideration of RB reuse and power control. They developed a Stackelberg game framework in which a CUE and a DUE pair is grouped to form a leader-follower pair.

	For multi-sharing D2D communication, Sun et al. \cite{GRA} aimed to maximize the number of permitted DUE pairs, given all transmission power values. They developed the Greedy Resource Allocation (GRA) algorithm. The key ideas of GRA is to form conflict graphs and to reuse RBs in the order of smallest degree first.

	Xu et al. \cite{Xu2014} jointly dealt with uplink resource allocation and power control by the iterative and auction-based algorithm called I-CAs. I-CAs considers the multi-sharing scenario in the sense that pre-dispatched ``packages'' of DUE pairs, rather than individual DUE pairs, reuse the RBs of CUEs. I-CAs highly relies on pre-dispatch of DUE pairs to packages, but how to obtain an optimal (or proper) pre-dispatch is a challenge since it is a combinatorial problem. I-CAs and GRA have a common restriction---a DUE pair can simultaneously reuse the RBs of up to one CUE. In \cite{Klugel2015}, Klugel and Kellerer studied whether it is feasible to satisfies all the SINR requirements under the power budgets, given a single CUE (or separate CUEs, each with a pre-dispatched package of DUE pairs) and given the SINR thresholds and power budgets.

%	In \cite{Klugel2015}, Klugel and Kellerer studied multi-sharing D2D communication from the ``feasibility'' point of view. Given a single CUE (or separate CUEs, each with a pre-dispatched package of DUE pairs) and given the SINR thresholds and power budgets on all user equipments (UEs), they investigated whether it is feasible to satisfies all the SINR requirements under the power budgets. They introduced two feasibility tests that are alternatives to each other. Test I calculates the maximum modulus eigenvalue of the corresponding Foschini matrix, computes the particular power vector (by a power-constrained Foschini-Miljanic power control algorithm \cite{Foschini1993}) that meets all SINR requirements, and checks if the particular power vector meets all power budgets. Test II exploits modified Rayleigh quotient; all SINR requirements can be satisfied under the power budgets if and only if test II outputs a value no greater than one.

	In \cite{Kao2015}, Ciou et al. proved NP-hardness for the multi-sharing resource allocation problem. This problem involves how to reuse of RBs that are allocated to CUEs, given the transmission power values on DUE pairs. In addition, they developed the GTM+ algorithm to get an efficient yet fast solution. The GTM+ algorithm exploits conflict graph and maximal weight independent set to improve system throughput while ensuring the minimum SINR requirements of CUEs and DUEs.

%	None of the above algorithms really studies throughput maximization of multi-sharing D2D communication by considering RB reuse and power control jointly. GRA \cite{GRA} and GTM+ \cite{Kao2015} do not deal with power control explicitly; I-CAs \cite{Xu2014} and the feasibility tests \cite{Klugel2015} highly rely on pre-dispatch of DUE pairs to packages, which remains a combinatorial problem. This motivates our study in considering RB reuse and power control jointly and in developing a new scheme that improves system throughput while being fast enough to support many DUEs in 5G mobile networks. We developed the \emph{m}aximum \emph{i}ndependent \emph{s}et based and \emph{S}tackelberg game based (MISS) algorithm. In MISS, RB reuse is accelerated by the help of maximum independent set and power control is done by the idea of Stackelberg game.

	None of the above algorithms really studies throughput maximization of multi-sharing D2D communication by considering RB reuse and power control jointly. GRA \cite{GRA} and GTM+ \cite{Kao2015} do not deal with power control explicitly; I-CAs \cite{Xu2014} highly relies on pre-dispatch of DUE pairs to packages, which remains a combinatorial problem. This motivates our study in considering RB reuse and power control jointly and in developing a new scheme that improves system throughput while being fast enough to support many DUEs in 5G mobile networks. We developed the \emph{m}aximum \emph{i}ndependent \emph{s}et based and \emph{S}tackelberg game based (MISS) algorithm. In MISS, RB reuse is accelerated by the help of maximum independent set and power control is done by the idea of Stackelberg game.

	The remainder of this paper is organized as follows. Section \ref{sec:system_model} describes the system model. We introduce in Section \ref{sec:Stackleberg_power_control} our Stackelberg-game-based power control method, which is designed for multi-sharing D2D communication. Section \ref{sec:algorithm_description} presents the MISS algorithm we propose. The performance evaluation and comparison are shown in Section \ref{sec:performance_evaluation}. We present some concluding remarks in Section \ref{sec:conclusion}.

\section{System Model}	\label{sec:system_model}

\begin{figure}[h]
	\begin{center}
		\includegraphics[width=0.9\hsize]{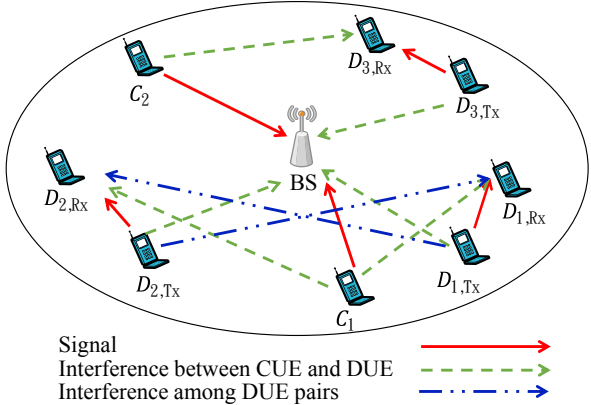}
		\caption{CUEs share RBs with DUEs ($C_1$ shares with $D_1$ and $D_2$; $C_2$ shares with $D_3$ in this figure), causing mutual interference.}
		\label{fig:D2D_reuse_UL}
	\end{center}
\end{figure}

	Although our method also applies to the downlink case, this paper focuses on the uplink case for exposition purpose. Same as many related papers, we consider a cell\footnote{Inter-cell interference is not considered because according to \cite{Xu2012} it can be managed efficiently with power control or resource scheduling. Although our algorithm can apply to the inter-cell case with some modification, for simplicity of exposition, this paper focuses on one cell at a time.} with $M$ CUEs and $N$ DUE pairs, as shown in Fig. \ref{fig:D2D_reuse_UL}. Within the cell, RBs are allocated disjointly/orthogonally among CUEs and these CUEs can share their pre-allocated \emph{uplink} RBs with DUE pairs. The CUEs and DUEs are denoted by $C_1, C_2, \dotsc, C_M$ and $D_1, D_2, \dotsc, D_N$, respectively. For simplicity of exposition, when there is no ambiguity, $c$ and $d$ are also used to denote $C_c$ and $D_d$, respectively. We denote the sender side of the DUE pair $d$ by $D_{d,\text{Tx}}$ and the receiver side by $D_{d,\text{Rx}}$.

	Depending on the numbers of allocated RBs, different CUEs can have either same or distinct bandwidths. For any certain CUE, say $c$, we denote its allocated bandwidth by $W_c$, its transmission power by $P_c$, and the noise power by $\sigma_c^2$. To make the system model as general as possible for the need of future 5G mobile networks, different CUEs are allowed to have different bandwidth (or equivalently, different numbers of allocated RBs), transmission power, and/or noise power.

	The serving base station (BS) is assumed to have the perfect channel state information of all communication channels and interference channels. The impact of major wireless channel impairments such as path loss, shadowing, and multipath fading can be incorporated into the channel gain. And we use $G_{\text{Tx,Rx}}$ to denote the channel gain from the transmitter Tx to the receiver Rx, where the transmitter and receiver can be a CUE $c$, a DUE pair $d$, and/or the serving base station $B$.

	%For complexity and cost reasons, any DUE pair is not allowed to simultaneously reuse two or more CUEs' RBs.

	To specify CUE-DUE relationship, $\Delta_c$ is defined as the set of DUE pairs that reuse the RBs allocated to CUE $c$. For the DUE pair $d$, we denote the transmission power at the sender by $P_d$, and the noise power at the receiver by $\sigma_d^2$. $P_d$ is often constrained within a range; that is, $P_{\min} \le P_d \le P_{\max}$.

	Note that because the sets $\{ \Delta_c \}$ specify RB reuse and the sets $\{ P_d \}$ specify power control, the values of $\Delta_c$ and $P_d$, for all $c \in \{ 1, 2, \dots, M \}$ and for all $d \in \{ 1, 2, \dots, N \}$, will be determined by the joint RB reuse and power control algorithm discussed in Section \ref{sec:algorithm_description}.

	All CUEs can have minimum SINR requirements. Each CUE, say $c$, can share its RBs with a set of DUEs if its SINR requirement is still satisfied. That is, the received SINR $\gamma_c$ must be beyond the SINR threshold $\gamma_c^t$:
\begin{equation}
	\gamma_c = \frac{P_c G_{c,B}}{\sum_{d \in \Delta_c} P_d G_{d,B} + \sigma_c^2}  \geq  \gamma_c^t	\label{eq:CUE_SINR_constraint}
\end{equation}
where $B$, $c$, and $d$ denotes the serving base station, CUE $c$, and DUE pair $d$, respectively; $G_{\text{Tx,Rx}}$ is the channel gain from the transmitter Tx to the receiver Rx.

	Similarly, each DUE pair, say $d$, also has its minimum SINR requirement. $d$ can reuse $c$'s RBs only if the received SINR $\gamma_d$ can exceed the SINR threshold $\gamma_d^t$:
\begin{multline}
	\gamma_d = \frac{P_d G_{d,d}}{P_{c} G_{c,d} + \sum_{d' \in \Delta_{c} - \{d\}} P_{d'} G_{d',d} + \sigma_d^2}  \geq  \gamma_{d}^t,  \\
		  \forall c: d \in \Delta_c	\label{eq:DUE_SINR_constraint}
\end{multline}

\section{Stackelberg-Game-Based Power Control} \label{sec:Stackleberg_power_control}

	This section outlines the Stackelberg-game-based power control method, which is a part of the MISS algorithm described later in Section \ref{sec:algorithm_description}. Roughly speaking, we extend Wang's idea proposed in \cite{Wang2013} to power control of multi-sharing D2D communication: In a general Stackelberg game with a leader and follower, the follower decides its best quantity, based on the price the leader offers, such that the follower's utility is maximized; the leader, who knows the follower's quantity function of the price variable, charges a fee/price for the follower so as to maximize the leader's utility. In this power control issue, the leader is a CUE; the follower is a DUE pair that wants to reuse the CUE's RBs; the CUE decides the price, which is a dummy variable; based on the price, the DUE pair decides its transmission power.

%	This section outlines the Stackelberg-game-based power control method, which is a part of the MISS algorithm described later in Section \ref{sec:algorithm_description}. Roughly speaking, we extend Wang's idea proposed in \cite{Wang2013} to power control of multi-sharing D2D communication: We treat this power control issue as a Stackelberg game, in which the CUE decides the price, which is a dummy variable and the DUE pair decides the quantity, which is its transmission power, based on the price.

	Consider a certain CUE, say $c$. Assume that all DUE pairs in the set $\Delta_c$ has been pre-determined to reuse $c$'s RBs and for each $d' \in \Delta_c$, its transmission power $P_{d'}$ has also been pre-determined. Now an extra DUE pair, say $d$, also wants to reuse $c$'s RBs. And our goal is to solve the Stackelberg game in order to obtain the best price $\alpha_c^*$ and the best transmission power $P_d^*$ such that $c$'s utility and $d$'s utility are both maximized.

	For DUE pair $d$, the utility function $U_d(\alpha_c, P_d)$ is defined as its throughput subtracted by the payment $d$ pays for reusing the RBs, which can be expressed as:
\begin{align}
	U_d(\alpha_c,P_d) =
		\log_2 \left( 1 + \frac{P_d G_{d,d}}{P_c G_{c,d} + \Phi} \right)
		- \alpha_c P_d G_{d,B}	\label{eq:DUE_utility function}
\end{align}
where $\Phi$ is defined as $\sum_{d' \in \Delta_{c}} P_{d'} G_{d',d} + \sigma_d^2$. The payment $d$ pays is set to be the price $\alpha_c$ multiplied by the interference $d$ imposes on the receiver side of $c$.

	For CUE $c$, its utility function is defined as its own throughput added by the revenue $c$ earns from $d$:
\begin{align}
	U_c(\alpha_c,P_d) =
		\log_2 \left( 1 + \frac{P_c G_{c,B}}{P_d G_{d,B} + \Omega} \right)
		+ \beta \alpha_c P_d G_{d,B}	\label{eq:CUE_utility function}
\end{align}
where $\Omega$ is defined as $\sum_{d' \in \Delta_c} P_{d'} G_{d',B} + \sigma_c^2$ and $\beta$ is a constant ratio of the revenue $c$ earns to the payment $d$ pays.

%	We use the backward induction to solve the Stackelberg game. We first do the follower analysis, which maximizes $d$'s utility. In the follower analysis, $\alpha_c$ is pretended as a constant and we obtain the best transmission power as a function of $\alpha_c$, which is denoted by $P_d^*(\alpha_c)$. Then we feed $P_d^*(\alpha_c)$ into $c$'s utility function and do the leader analysis to maximize $c$'s utility. After the leader analysis, we obtain the optimal price $\alpha_c^*$. With the value $\alpha_c^*$, the best transmission power $P_d^*$ can be easily computed by $P_d^* = P_d^*(\alpha_c^*)$.

	Due to the space problem, we omit the analysis and only present the result: The optimal price $\alpha_c^*$ takes only on one of the six values $\{ \alpha_{c,1}, \alpha_{c,2}, \alpha_{c,3}, \alpha_{c,4}, \alpha_{c,\min}, \alpha_{c,\max} \}$, where
\begin{align}
\begin{aligned}
&	\alpha_{c,1} = \frac{B}{\beta \Omega} - \frac{B}{A} \\
&	\alpha_{c,2} = \frac{B}{A} - \frac{B}{(A+\Omega) \beta} \\
&	\alpha_{c,3} = \frac{-B(A+2C)-\sqrt{D}}{2C(A+C)} \\
&	\alpha_{c,4} = \frac{-B(A+2C)+\sqrt{D}}{2C(A+C)} \\
&	\alpha_{c,\min} = \frac{B}{P_{\max} G_{d,B} + \Omega - C} \\
&	\alpha_{c,\max} = \frac{B}{P_{\min} G_{d,B} + \Omega - C}
\end{aligned}
\label{eq:possible_optimal_prices_alpha_c}
\end{align}
and
\begin{align*}
&	A = P_c G_{c,B} \\
&	B = \frac{1}{\ln 2} \\
&	C = -\frac{G_{d,B}}{G_{d,d}} (P_c G_{c,d}+\Phi) + \Omega \\
&	D = A B^2 \left( A + 4C (A+C) \frac{1}{(\Omega-C) \beta} \right)
\end{align*}
With these six possible $\alpha_c^*$, the corresponding best transmission power can be easily computed by:
\begin{align}
	P_d^* = \begin{cases}
		\hat{P_d}	& \text{if } P_{\min} \le \hat{P_d} \le P_{\max}  \\
		P_{\min}	& \text{if } \hat{P_d} < P_{\min}  \\
		P_{\max}	& \text{if } \hat{P_d} > P_{\max}
\end{cases}   \label{eq:best_DUE_power_Pd*}
\end{align}
where
\[  \hat{P_d} = \frac{1}{\alpha_c^* G_{d,B} \ln 2} - \frac{P_c G_{c,d} + \Phi}{G_{d,d}}  \]

	What our Stackelberg-game-based power control method needs to do is to find, among the six $(\alpha_c, P_d)$ points, the one that makes $U_c(\alpha_c, P_d)$ largest. This gives the best transmission power value $P_d^*$. Since it takes six operations, the Stackelberg-game-based power control has a complexity of $\mathcal{O}(1)$.

\subsection{Follower Analysis}

	We first do the follower analysis, in which $\alpha_C$ is pretended as a constant. DUE pair $d$ aims to maximize its utility by determining its best transmission power. Obviously, $d$'s utility function is concave with respect to $P_d$ and thus the maximum value exists. The maximum point can be found by taking the partial derivation:
\begin{equation*}
	\pd{U_d}{P_d}=\frac{1}{\ln 2}\frac{G_{d,d}}{P_d G_{d,d}+P_c G_{c,d}+\Phi}-\alpha_c G_{d,B}
\end{equation*}
The above derivation equals zero when the transmission power on the sender side of $d$ equals
\begin{align} \label{eq:DUE_optpower}
	\hat{P_d} = \frac{1}{\alpha_c G_{d,B} \ln 2} - \frac{P_c G_{c,d}+\Phi}{G_{d,d}}
\end{align}
Note that because of the constraint $P_{\min} \leq P_d^* \leq P_{\max}$, given $\alpha_c$, the best transmission power $P_d^*(\alpha_c)$ is searched in $\{ P_{\min}, P_{\max}, \hat{P_d} \}$.

\subsection{Leader Analysis}

	By backward induction, the leader knows ex ante that the follower will react to its price by searching in $\{ P_{\min}, P_{\max}, \hat{P_d} \}$. If the price that the leader sets too low, the follower will only buy $P_{max}$, if the price that the leader sets too high, the follower will choose to buy $P_{min}$ or not to use. Thus, without loss of generality, the leader will set a price as $P_{min}\leq P_d \leq P_{max}$. By substituting \eqref{eq:DUE_optpower} into \eqref{eq:CUE_utility function}, we get
\begin{multline}  \label{complicated CUE function}
	U_c(\alpha_c) =
		\frac{\beta}{\ln 2}
		-\alpha_c\beta*\frac{G_{d,B}}{G_{d,d}}*(P_c G_{c,d}+\Phi)  \\
		+\log_2 \left( 1+\frac{P_c G_{c,B}}{\frac{1}{\alpha_c \ln 2}-\frac{G_{d,B}}{G_{d,d}}*(P_c G_{c,d}+\Phi)+\Omega} \right)
\end{multline}

	There is a tradeoff between the gain from the leader itself and the gain from the follower. When the leader raises the price, the follower will buy less power. Therefore, the leader has to find out the optimal price to maximize its utility.

	Let $A = P_c G_{c,B}, B = \frac{1}{\ln 2}, C = -\frac{G_{d,B}}{G_{d,d}}*(P_c G_{c,d}+\Phi)+\Omega$, we can rewrite (\ref{complicated CUE function}) as:
\begin{equation}  \label{reduced CUE function}
	U_c(\alpha_c) = \log_2( 1 + \frac{A\alpha_c}{C \alpha_c + B}) + B \beta - \alpha_c \beta (\Omega - C).
\end{equation}

	We can get the optimal price by analyzing the first-order derivative of \eqref{reduced CUE function}:
\begin{equation}  \label{first derivative of CUE function}
	\fd{U_c}{\alpha_c} = \frac{A B^2}{(C \alpha_C + B)((A + C) \alpha_C + B)} - \beta(\Omega-C)
\end{equation}
Denote the denominator of the first term on the right-hand side of \eqref{first derivative of CUE function} by $f(\alpha_c) = (C \alpha_c + B) ((A + C) \alpha_c + B)$. To get the optimal price that maximizes $c$'s utility, we consider (\ref{first derivative of CUE function}) the following five situations:

\begin{description}  \parskip 0pt \parsep 0pt
	\item[1. $C=0$.] \hfill \\
		$\fd{U_c}{\alpha_c} = \frac{A B^2}{B (A \alpha_c + B)}- \beta \Omega$. The derivative equals zero at $\alpha_c = \frac{B}{\beta \Omega} - \frac{B}{A}$.

		We define $\alpha_{c,1} = \frac{B}{\beta \Omega} - \frac{B}{A}$, because the second order derivative is $\frac{d^2 U_c}{d {\alpha_c}^2} = - B(\frac{A}{A \alpha_c + B})^2 < 0$. So optimal $\alpha_c^*$ can be searched in $\{\alpha_{c,1}, \alpha_{c,\min}, \alpha_{c,\max}\}$, where $\alpha_{c,\min} = \frac{B}{P_{\max} G_{d,B} + \Omega - C}$ and $\alpha_{c,\max} = \frac{B}{P_{\min} G_{d,B} + \Omega - C}$.
	
		Note that when $C<0$, $\alpha_{c,\max}<-\frac{B}{C}$.

	\item[2. $C<0$ and $A+C=0$.] \hfill \\
		$\fd{U_c}{\alpha_c} = \frac{A B^2}{(-A \alpha_c + B)B} - \beta (A + \Omega)$. The derivative equals zero at $\alpha_c = \frac{B}{A} - \frac{B}{(A + \Omega) \beta}$. Define $\alpha_{c,2} = \frac{B}{A} - \frac{B}{(A+\Omega) \beta}$. In this case, $U'_c({\alpha_c})$ is also a concave function because $\frac{d^2 U_c}{d {\alpha_c}^2} = \frac{A^2 B}{(-A \alpha_c + B)^2} < 0$. So the optimal $\alpha_c$ can also be searched in $\{ \alpha_{c,2}, \alpha_{c,\min}, \alpha_{c,\max}\}$. \footnote{In the case when $A + C = 0$, either $\alpha_{c,\min}$ or $\alpha_{c,\max}$ results in a higher utility than $\alpha_{c,2}$.}

	\item[3. $C>0$.] \hfill \\
		In this case, the two roots of the quadratic function $f(\alpha_c)$, which are $-\frac{B}{C}$ and $-\frac{B}{A+C}$, are both negative. This implies that $f(\alpha_c)$ is positive and monotonically increasing in the right half-plane. So the derivative of utility, $U'_c(\alpha_c)$, is monotonically decreasing in the right half-plane. Note that a charging price $\alpha_c$ is a positive number no smaller than $\alpha_{min}$ and therefore we pay attention only to the right half-plane. Consider the value of $U'_c$'s derivative. If $U'_c(\alpha_{min})\leq 0$, the optimal price is $\alpha_{min}$. If $U'_c(\alpha_{min})>0$, by solving the quadratic equation $f(\alpha_c) - \frac{A B^2}{\beta (\Omega - C)} = 0$, we obtain that the two roots of $U'_c(\alpha_c)$ are $\alpha_{c,3} = \frac{-B(A+2C)-\sqrt{D}}{2C(A+C)}$, $\alpha_{c,4} = \frac{-B(A+2C)+\sqrt{D}}{2C(A+C)}$, where $D = A B^2 \left( A + 4C (A + C) \frac{1}{(\Omega - C) \beta} \right)$. The smaller root of $U'_c(\alpha_c)$ is out of our interest because the smaller root is negative. Only the larger root could be a valid charging price $\alpha_c$. Therefore, the maximum of $U_c(\alpha_c)$ happens at the larger root.

	\item[4. $C<0$ and $A+C>0$.] \hfill \\
		In this case, $f(\alpha_c)$ has one positive root $-\frac{B}{C}$ and one negative root $-\frac{B}{A+C}$; $f(\alpha_c)$ is positive in $(-\frac{B}{A+C}, -\frac{B}{C})$, is zero at $-\frac{B}{A+C}$ and $-\frac{B}{C}$, and is negative elsewhere. Under such circumstance, if $U'_c(\alpha_c) \geq 0 \; \forall \alpha_c$ in the feasible region $[\alpha_{c,\min}, \alpha_{c,\max}]$, then the maximum point is $\alpha_{c,\max}$.

		Otherwise, considering that $f(\alpha_c)$ is a concave quadratic function, where $-\frac{B}{C} >0$ and $-\frac{B}{A+C}<0$, and $-\frac{B}{A+C}<0<\alpha_{c,\min}\leq\alpha_{c} \leq \alpha_{c,\max} < -\frac{B}{C}$, we know that as $\alpha_c$ increases, $U'_c(\alpha_c)$ will first decreases and then increases sequentially. Thus, the maximum point can be searched in $\{\alpha_{c,3}, \alpha_{c,\min}, \alpha_{c,\max} \}$.

	\item[5. $C<0$ and $A+C<0$.] \hfill \\
		Similar to case 3, but $-\frac{B}{A+C}>-\frac{B}{C}>\alpha_{c,\max} >\alpha_{c,\min}>0$, which means within the feasible region, $f(\alpha_c)>0,$ and $f(\alpha_c)$ is monotonically decreasing with $\alpha_c$, $U'_c(\alpha_c)$ is monotonically increasing with $\alpha_c$. Thus, the optimal price $\alpha_c^*$ is either $\alpha_{c,\min}$ or $\alpha_{c,\max}$.

\end{description}

\section{The MISS Algorithm}	\label{sec:algorithm_description}

	For joint RB reuse and power control of multi-sharing D2D communication, we devise the \emph{m}aximum \emph{i}ndependent \emph{s}et based and \emph{S}tackelberg game based (MISS) algorithm. MISS run iteratively until all CUEs are marked. In each iteration, MISS decides which DUE pairs and at what transmission power to reuse the RBs pre-allocated to a certain CUE. The RB reuse part of MISS exploits maximum independent set, resulting in a small subset of DUE pairs that need to adjust transmission power. The power control part of MISS is accelerated by employing the aforementioned Stackelberg-game-based method with a $\mathcal{O}(1)$ time complexity. As shown in the pseudo code, MISS consists of one-time initialization and a number of iterations which can be further divided into three major steps.

	\textbf{Initialization:} At the beginning, all groups are unmarked and every DUE pair joins the group that maximizes the \emph{sheer rate}. Each group is owned by an exclusive CUE; so, group and CUE can be thought as aliases of each other in this paper. The sheer rate of group/CUE $c$ and DUE pair $d$
\begin{equation}	\label{eq:sheer_rate}
	r(c,d) =	W_c \log_2 (1 + \tfrac{P_c G_{c,B}}{\sigma_c^2 + P_d^* G_{d,B}})
				+ W_c \log_2 (1 + \tfrac{P_d^* G_{d,d}}{\sigma_d^2 + P_c G_{c,d}})
\end{equation}
is defined as the sum of $c$'s and $d$'s throughputs, assuming that no other DUE pair reuses $c$'s RBs. $P_d^*$ is computed by the power control method described in Section \ref{sec:Stackleberg_power_control}.

	\textbf{Step 1:} The goal of this step is for the largest unmarked group, say group $c$, to find out the \emph{proper DUE pairs}. The set of all proper DUE pairs is denoted by $\Lambda_c$.

	$\Lambda_c$ is found by taking the maximum independent set\footnote{Because the maximum independent set problem is NP-hard, we use the heuristic algorithm in \cite{MWIS} to obtain a maximal independent set instead. The heuristic is of time complexity $\mathcal{O}(n^3)$.} of a conflict graph $\mathcal{G}$. The vertices in the conflict graph correspond to the DUE pairs corresponding to all DUE pairs that have not been granted to reuse any RB. For any two vertices, a connecting edge is added to the conflict graph if their distance is smaller than a predetermined threshold.

	For speed-up purpose, improper DUE pairs are ignored; only the proper DUE pairs $\Lambda_c$, which are DUE pairs in the maximum independent set, are considered (but not guaranteed) to reuse $c$'s RBs.

	\textbf{Step 2:} Not every proper DUE pair  will eventually reuse $c$'s RBs; instead, Step 2 picks a subset of proper DUE pairs that reuse $c$'s RBs, denoted by $\Delta_c$, incrementally by a best-fit strategy consisting of $L$ rounds.

	In each round, all DUE pairs in $\Delta_c$ are first checked to ensure that their SINR requirements are not violated because of adding a new DUE pair in the previous round. The DUE pairs failed to meet their SINR requirements are moved to $\Lambda_c$ for another trial. After that, the DUE pair in $\Lambda_c$ that has the highest \emph{pairwise throughput} is picked and moved to $\Delta_c$ from $\Lambda_c$. The pairwise throughput of CUE $c$ and DUE pair $d$ is defined as the sum of $c$'s throughput and $d$'s throughput if $c$'s and $d$'s SINR requirements are both met; otherwise, the pairwise throughput is set to zero:
\begin{align}  \label{eq:pairwise_throughput}
\lambda(c,d) = \left\{
\begin{aligned}
		& \log_2 \left( 1  + \tfrac{P_c G_{c,B}}{P_d^* G_{d,B} + \Omega} \right)	 +  \log_2 \left( 1 + \tfrac{P_d^* G_{d,d}}{P_c G_{c,d} + \Phi} \right), 	\\
		& \qquad\qquad  \text{ if } \tfrac{P_c G_{c,B}}{P_d^* G_{d,B} + \Omega}  \ge \gamma_c^t \text{ and } \tfrac{P_d^* G_{d,d}}{P_c G_{c,d} + \Phi} \ge \gamma_d^t  \\
		& 0, \qquad\quad\:\, \text{if } \tfrac{P_c G_{c,B}}{P_d^* G_{d,B} + \Omega}  < \gamma_c^t \text{ or  } \tfrac{P_d^* G_{d,d}}{P_c G_{c,d} + \Phi} < \gamma_d^t
\end{aligned} \right.
\end{align}
where $\Omega$ is defined as $\sum_{d' \in \Delta_c} P_{d'} G_{d',B} + \sigma_c^2$ and $\Phi$ is defined as $\sum_{d' \in \Delta_{c}} P_{d'} G_{d',d} + \sigma_d^2$. $P_d^*$ is computed by using the Stackelberg-game-based power control method described in Section \ref{sec:Stackleberg_power_control}.

	After that, all of the proper DUE pairs in $\Delta_c$ are granted to reuse $c$'s RBs at their own transmission power levels computed by using our Stackelberg-game-based power control method.

	\textbf{Step 3:} After the best-fit strategy in Step 2, there might exist proper DUE pairs that are not picked to reuse $c$'s RBs. Step 3 lets such DUE pairs join the unmarked groups (excluding group $c$) that give them highest sheer rates. After that, the proper DUE pairs granted to reuse $c$'s RBs are removed from the conflict graph $\mathcal{G}$. And group $c$, which used to be the largest unmarked group, is set to be marked.

\begin{algorithm}
%\SetAlgoNoLine
%\SetCommentSty{textrm}
%\SetCommentSty{textit}
%\LinesNumbered
%\SetAlFnt{\small}

\footnotesize
	% The simplest way would be to add \small, \footnotesize, \scriptsize, or \tiny immediately after the beginning of the algorithm environment. This changes the size of everything inside the environment (except for the caption), and nowhere else.
	% This will apparently leave the size of the line-numbers and of the caption unchanged. In order to change the size of the line-numbers for the algorithm, add the command \algsetup{linenosize=<size>} just inside the algorithm environment, where <size> is again something like \small, \footnotesize, etc. Note that the default is to have line-numbers somewhat smaller than the surrounding text: if you want the appearance to be somewhat similar to algorithms elsewhere in your document, you should also choose a smaller size for the line-numbers than the algorithm body.

	\caption{MISS}
	\label{algo:MISS}

\myAlgo{MISS}{

	\KwIn{$M$ CUEs and $N$ DUE pairs.}

	\KwOut{RB reuse results $\{ \Delta_1, \Delta_2, \dotsc, \Delta_M \}$ and transmit power results $\{ P_1^*, P_2^*, \dotsc, P_N^* \}$.}

%\hangindent=1em

	\tcp{Initialization.}

	$U  \gets  \{ 1,2,\dotsc,M \}$.  \tcp{$U$ is the set of unmarked groups/CUEs.}

	Set $\Gamma_1, \Gamma_2, \dotsc, \Gamma_M$ and $\Delta_1, \Delta_2, \dotsc, \Delta_M$ to be empty sets.  \tcp{$\Gamma_c$ is the set of DUEs that joins group $c$.}

	\ForEach {$d \in \{ 1, 2, \dots, N \}$}
	{

		$c^*  \gets  $ WhoGivesMaxSheerRate($d$, $U$).

		$\Gamma_{c^*}  \gets  \Gamma_{c^*} \cup  \{d\}$.  \tcp{That is, $D_d$ joins group $c^*$.}

	}

	Form the conflict graph $\mathcal{G}$ for all DUE pairs.

	\tcp{The main body (consisting of iterations) starts here.}
	
	%\While {$U \ne \emptyset$ and at least one element in $\{ \Gamma_m : m \in U \}$ is non-empty}
	\While {$U \ne \emptyset$}
    {

		%Form the conflict graph $G_{c}$ for the largest group $c$ in $U$.  \tcp{$c  \gets  \argmax_{c' \in U} | \Gamma_{c'} |$.}

		$\Lambda_{c}  \gets$  the maximum independent set of $\mathcal{G}$.

		\tcp{The best-fit strategy consisting of $L$ rounds starts here.}

		\ForEach {$l \in \{ 1,2,\dotsc, L \} $}
		{

			\ForEach {$d \in \Delta_c$}
			{
				\tcp{Check if DUE pair $d$ does not meet its SINR requirement.}

				$P_d^*  \gets$ StackelbergPowerControl($d$, $c$, $\Delta_c$).

				\If {$\frac{P_d^* G_{d,d}}{P_c G_{c,d} + \Phi} < \gamma_d^t$}
				{
					Move $d$ from $\Delta_c$ to $\Lambda_c$.
				}

			}

			$(d, P_d^*, \text{PairwiseThru})  \gets$ MaxPairwiseThru($\Lambda_c, c, \Delta_c$).  \tcp{Find the DUE pair that has the highest pairwise throughput.}

			\If { $\text{PairwiseThru} > 0$ }
			{
				Update $\Lambda_c$ and $\Delta_c$. \tcp{Move $D_d$ from $\Lambda_c$ to $\Delta_c$.}
			}

		}

		\ForEach {$d \in \Gamma_c - \Delta_c$}
		{

			$c^*  \gets  $ WhoGivesMaxSheerRate($d$, $U - \{c\}$).

			$\Gamma_{c^*}  \gets  \Gamma_{c^*} \cup \{d\}$.  \tcp{$D_d$ joins group $c^*$.}
			
		}

		Remove $\Delta_c$ from the conflict graph $\mathcal{G}$.
		
		$U  \gets  U - \{ c \}$. \tcp{Make group $c$ marked.}

	}

}

\myFn{WhoGivesMaxSheerRate($d$, $C$)}{

	\tcp{Return $c^* = \argmax_{c \in C} \, \{ r(c,d) : \frac{P_c G_{c,B}}{P_d^* G_{d,B} + \Omega}  \ge \gamma_c^t \}$.}

	MaxValue  $\gets$  0.

	$c^*  \gets  0$.

	\ForEach {$c \in C$}
	{
		$P_d^*  \gets$ StackelbergPowerControl($d$, $c$, $\emptyset$).

		\If { $\frac{P_c G_{c,B}}{P_d^* G_{d,B} + \Omega}  \ge \gamma_c^t$ }
		{
			\If { $r(c,d) > $ MaxValue }
			{
				MaxValue  $\gets$  $r(c,d)$.

				$c^*  \gets  c$.
			}
		}
	}

	Return $c^*$.

}

\myFn{MaxPairwiseThru($\Lambda_c, c, \Delta_c$)}{

	%\tcp{Return $d^* = \argmax_{d \in \Lambda_c} \, \{ \lambda(c,d) : \gamma_c \ge \gamma_c^t, \gamma_d \ge \gamma_d^t \}$.}

	\tcp{Find the DUE pair that has the highest pairwise throughput. Return the DUE pair, its best transmission power, and the value of its pairwise throughput.}

	MaxValue  $\gets$  0.

	$d^*  \gets  0$.

	\ForEach {$d \in \Lambda_c$}
	{

		$P_d^*  \gets$ StackelbergPowerControl($d$, $c$, $\Delta_c$).

		\If { $\frac{P_c G_{c,B}}{P_d^* G_{d,B} + \Omega}  \ge \gamma_c^t$ and $\frac{P_d^* G_{d,d}}{P_c G_{c,d} + \Phi} \ge \gamma_d^t$ }
		{
			\If { $\lambda(c,d) > $ MaxValue }
			{
				MaxValue  $\gets$  $\lambda(c,d)$.

				$d^*  \gets  d$.
			}
		}
	}

	Return $(d^*, P_d^*, \text{MaxValue})$.

}

\myFn{StackelbergPowerControl($d$, $c$, $\Delta_c$)}{

	\tcp{Return the best transmission power $P_d^*$ determined by the Stackelberg-game-based power control.}

\hangindent=1em
	Calculate, by \eqref{eq:possible_optimal_prices_alpha_c}, the six possible optimal prices---$\alpha_{c,1}$, $\alpha_{c,2}$, $\alpha_{c,3}$, $\alpha_{c,4}$, $\alpha_{c,\min}$, and $\alpha_{c,\max}$.

\hangindent=1em
	Calculate, by \eqref{eq:best_DUE_power_Pd*}, the best transmission power values $\{ P_{d,1}, P_{d,2}, P_{d,3}, P_{d,4}, P_{\min}, P_{\max} \}$, each corresponding to one element in $\{ \alpha_{c,1}, \alpha_{c,2}, \alpha_{c,3}, \alpha_{c,4}, \alpha_{c,\min}, \alpha_{c,\max} \}$.

\hangindent=1em
	Calculate $c$'s utility values, by \eqref{eq:CUE_utility function}, at the six $(\alpha_c, P_d)$ points.

\hangindent=1em
	Return the transmission power corresponding to the point at which $c$'s utility is largest.

}

\end{algorithm}

\section{Performance Evaluation}	\label{sec:performance_evaluation}

	We evaluate the performance of our proposed MISS algorithm by using our in-house simulator and compare its performance with three existing algorithms---GTM+ \cite{Kao2015}, GRA \cite{GRA}, and ORA \cite{ORA}. MISS, GTM+, and GRA are designed for multi-sharing D2D communication, whereas ORA is designed for single-sharing D2D communication. %From the power control aspect, ORA does not really treat CUEs and DUEs differently, while MISS considers a primary-secondary relationship---CUEs are primary users and DUE pairs are secondary users. Given the transmission power values of primary users, MISS can adjust transmission power of secondary users only (such that secondary users benefit by reuse of RBs without causing too much interference on primary users). On the other hand, ORA essentially regards CUEs and DUEs as peers and adjusts transmission power of both. GTM+ and GRA have no power control mechanism; they assume that transmission power control values are given.

	The simulation is set as follows. The number of CUEs $M$ varies from 40 to 110. 110 is the number of RBs a 20MHz LTE/LTE-A system can have in theory. The ratio of DUE pairs to CUEs is set to four. For fair comparison with the the existing algorithms, all CUEs are set to have the same SINR threshold and noise spectral density (although for MISS it does not have to). All DUE pairs are set to have equal SINR threshold and noise spectral density, too. Each RB is allocated to a CUE and each CUE is allocated a RB in the simulation. These settings are for fair comparison purpose; otherwise, some algorithms cannot be applied. For the algorithms which have power control capability, the transmission power range is between 0 Watt and 23 dBm. For those which have no power control, each CUE has a fixed transmission power of 23 dBm and each DUE has a fixed transmission power of 10 dBm.

	All CUEs and DUEs are randomly distributed in a single cell with the serving BS at the center. Most parameters are set according to \cite{GRA}; some of them are listed in Table \ref{table:simulation_parameter}. All results are averaged over at least 100 instances. Important performance metrics include system throughput (which is normalized to have a unit of bit/s/Hz) and DUEs' total transmission power. We also evaluate the percentage of permitted DUE pairs and the running time each algorithm takes.

\begin{table}[t]
	\caption{Simulation parameters.}
	\label{table:simulation_parameter}
	
	\centering
    \begin{tabular}[c]{|l|l|}
		\hline
		\textbf {Parameters} & \textbf {Value} \\
		\hline
		CUE transmission power (fixed) & 23 dBm \\
		\hline
		CUE transmission power (adjustable) & 0 Watt to 23 dBm \\
		\hline
		DUE transmission power & 10 dBm \\
		\hline
		DUE transmission power (adjustable) & 0 Watt to 23 dBm \\
		\hline
		Radius of BS coverage & 500 m \\
		\hline
		Noise spectral density & -174 dBm/Hz \\
		\hline
		Path loss model for CUE and DUE & $128.1 + 37.6 \log_{10}(d \text{ [km]})$ \\
		\hline
		Path loss model for DUE pairs & $148 + 40 \log_{10}(d \text{ [km]})$ \\
		\hline
%		Noise power $(\sigma_m^2, \sigma_n^2)$ & -121.45 dBm \\
%		\hline
		CUE's SINR threshold & 7 \\
		\hline
		DUE's SINR threshold & 3 \\
		\hline
		The distance between each DUE pair & 15 m \\
		\hline
		Bandwidth per RB & 12 * 15kHz = 180 kHz \\
		\hline
    \end{tabular}
\end{table}

\begin{figure}[h]
    \begin{center}
        \includegraphics[width=0.8\hsize]{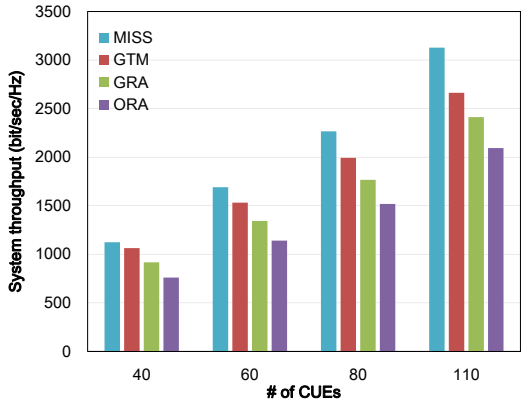}
        \caption{System throughput.}
        \label{fig:throughput__C1D4}
    \end{center}
\end{figure}

	In terms of system throughput\footnote{System throughput is the sum of the Shannon capacity values of all CUEs and DUEs that satisfy their own SINR requirements.}, MISS performs best in all cases, GTM+ is the second place, GRA is the third place, and ORA performs worst, as seen in Fig. \ref{fig:throughput__C1D4}. The major reason of the outperformance is because only MISS has both power control and multi-sharing capabilities: MISS's power control capability can reduce interference imposed on CUEs and DUEs, which cannot be done by GTM+ and GRA. MISS's multi-sharing capability allows multiple DUE pairs to reuse same RBs, which cannot be done by ORA.

\begin{figure}[h]
    \begin{center}
        \includegraphics[width=0.8\hsize]{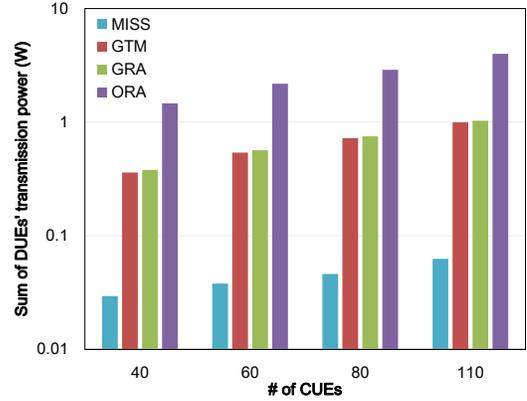}
        \caption{DUEs' total transmission power.}
        \label{fig:DUE_power__C1D4}
    \end{center}
\end{figure}

\begin{figure}[h]
    \begin{center}
        \includegraphics[width=0.8\hsize]{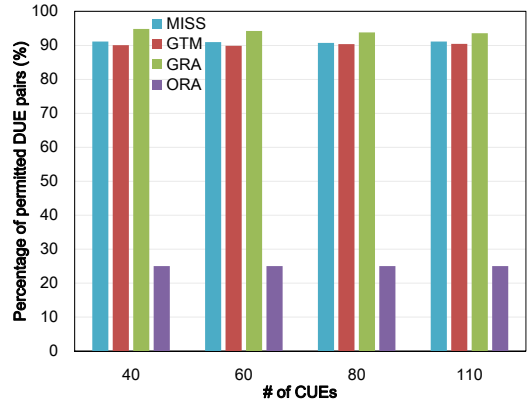}
        \caption{The percentage of permitted DUE pairs.}
        \label{fig:permitted_DUE_pairs}
    \end{center}
\end{figure}

	In terms of DUEs' total transmission power, a dramatic improvement by MISS is observed in Fig. \ref{fig:DUE_power__C1D4}. MISS results in a much smaller transmission power than other three algorithms. The reason behind is that MISS aims to both maximize throughput and minimize interference. These two objectives are converted into the first term (throughput maximization) and the second term (interference minimization) of DUE's utility function \eqref{eq:DUE_utility function}. On the other hand, ORA tends to increase CUEs' and DUEs' transmission power proportionally so as to overcome noise. The transmission power consumption values of GTM+ and GRA are somewhat close to each other because GTM+ and GRA have no power control capability and the percentage of permitted DUE pairs (as shown in Fig. \ref{fig:permitted_DUE_pairs}) becomes a dominant factor in total transmission power. Note that MISS performs drastically better than the other algorithms in terms of system throughput per power because compared with other three algorithms, MISS results in highest throughput at lowest transmission power.

	As shown in Fig. \ref{fig:permitted_DUE_pairs}, GRA permits more DUE pairs to reuse RBs than the other three algorithms do. MISS, which permits roughly 90\% of DUE pairs for RB reuse, is the second place. The third place is GTM+, which performs worse than MISS insignificantly. ORA performs worst; this is because ORA is not designed for multi-sharing D2D communication.

\begin{figure}[h]
    \begin{center}
        \includegraphics[width=0.8\hsize]{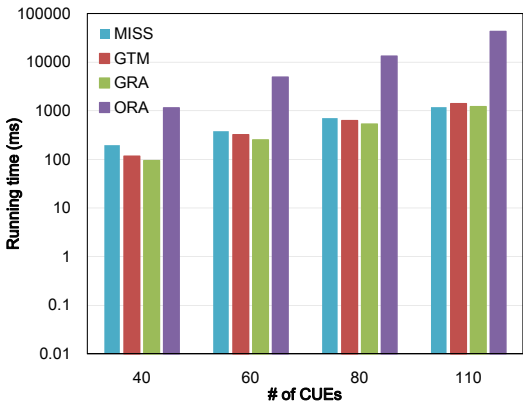}
        \caption{The running time each algorithm takes.}
        \label{fig:running_time}
    \end{center}
\end{figure}

	Besides the performance indices aforementioned, the running time each algorithm takes is presented in Fig. \ref{fig:running_time}. As one can observe, MISS is roughly as fast as GTM+ and GRA. In addition, MISS is one to two orders of magnitude faster than ORA in the simulation setup. Note that the running time results are obtained from executing Matlab code; in real systems implementing in C/C++ or with hardware acceleration, the MISS algorithm can easily complete in a much shorter time than what Fig. \ref{fig:running_time} shows.

\section{Conclusion}	\label{sec:conclusion}

	Next-generation mobile networks are expected to support higher numbers of simultaneously connected devices and to achieve higher system spectrum efficiency and lower power consumption. To achieve these goals, we have studied the multi-sharing device-to-device (D2D) communication with resource block reuse and power control jointly considered. We have proposed the MISS algorithm, which exploits and gets accelerated by the use of maximum independent set and Stackelberg game. Extensive simulation results show that compared to the three existing algorithms, MISS has outstanding performance in terms of transmission power consumption, system throughput, the percentage of permitted DUE pairs, and running time. In particular, MISS results in highest system throughput and lowest transmission power consumption.

\section*{Acknowledgment}

	This work was supported in part by the Ministry of Science and Technology (MOST), Taiwan, under grants no. MOST 103-2221-E-007-041-MY3, MOST 103-2218-E-007-022, MOST 104-3115-E-007-005, and MOST 105-2218-E-007-012.

\bibliography{IEEEabrv,ref}

\begin{thebibliography}{10}
\providecommand{\url}[1]{#1}
\csname url@samestyle\endcsname
\providecommand{\newblock}{\relax}
\providecommand{\bibinfo}[2]{#2}
\providecommand{\BIBentrySTDinterwordspacing}{\spaceskip=0pt\relax}
\providecommand{\BIBentryALTinterwordstretchfactor}{4}
\providecommand{\BIBentryALTinterwordspacing}{\spaceskip=\fontdimen2\font plus
\BIBentryALTinterwordstretchfactor\fontdimen3\font minus
  \fontdimen4\font\relax}
\providecommand{\BIBforeignlanguage}[2]{{%
\expandafter\ifx\csname l@#1\endcsname\relax
\typeout{** WARNING: IEEEtran.bst: No hyphenation pattern has been}%
\typeout{** loaded for the language `#1'. Using the pattern for}%
\typeout{** the default language instead.}%
\else
\language=\csname l@#1\endcsname
\fi
#2}}
\providecommand{\BIBdecl}{\relax}
\BIBdecl

\bibitem{key_parameters_for_5G}
\BIBentryALTinterwordspacing
H.-R. You. (2015, Mar.) Key parameters for {5G} mobile communications [{ITU-R
  WP 5D} standardization status]. KT. [Online]. Available:
  \url{http://www.netmanias.com/en/post/blog/7335/5g-kt/key-parameters-for-5g-mobile-communications-itu-r-wp-5d-standardization-status}
\BIBentrySTDinterwordspacing

\bibitem{Kao2015}
S.-A. Ciou, J.-C. Kao, C.~Y. Lee, and K.-Y. Chen, ``Multi-sharing resource
  allocation for device-to-device communication underlaying {5G} mobile
  networks,'' in \emph{IEEE Intl. Symposium on Personal, Indoor, and Mobile
  Radio Communications (PIMRC)}, Hong Kong, Aug.-Sep. 2015.

\bibitem{Doppler2009}
K.~Doppler, M.~Rinne, C.~Wijting, C.~B. Ribeiro, and K.~Hugl,
  ``Device-to-device communication as an underlay to {LTE-Advanced} networks,''
  \emph{{IEEE} Commun. Mag.}, vol.~47, no.~12, pp. 42--49, Dec. 2009.

\bibitem{Yu2011}
C.-H. Yu, K.~Doppler, C.~B. Ribeiro, and O.~Tirkkonen, ``Resource sharing
  optimization for device-to-device communication underlaying cellular
  networks,'' \emph{{IEEE} Trans. Commun.}, vol.~10, no.~8, pp. 2752--2763,
  Aug. 2011.

\bibitem{Han2012}
T.~Han, R.~Yin, Y.~Xu, and G.~Yu, ``Uplink channel reusing selection
  optimization for device-to-device communication underlaying cellular
  networks,'' in \emph{IEEE Intl. Symposium on Personal Indoor and Mobile Radio
  Communications (PIMRC)}, Sep. 2012.

\bibitem{ORA}
D.~Feng, L.~Lu, Y.~Yuan-Wu, G.~Li, G.~Feng, and S.~Li, ``Device-to-device
  communications underlaying cellular networks,'' \emph{{IEEE} Trans. Commun.},
  vol.~61, no.~8, pp. 3541--3551, Aug. 2013.

\bibitem{Wang2013}
F.~Wang, L.~Song, Z.~Han, Q.~Zhao, and X.~Wang, ``Joint scheduling and resource
  allocation for device-to-device underlay communication,'' in \emph{IEEE
  Wireless Communications and Networking Conference (WCNC)}, Shanghai, China,
  Apr. 2013.

\bibitem{GRA}
H.~Sun, M.~Sheng, X.~Wang, Y.~Zhang, J.~Liu, and K.~Wang, ``Resource allocation
  for maximizing the device-to-device communications underlaying {LTE-Advanced}
  networks,'' in \emph{IEEE/CIC International Conference on Communications in
  China (ICCC) - Workshops}, Aug. 2013.

\bibitem{Xu2014}
C.~Xu, L.~Song, D.~Zhu, and M.~Lei, ``Subcarrier and power optimization for
  device-to-device underlay communication using auction games,'' in \emph{IEEE
  International Conference on Communications (ICC)}, Jun. 2014.

\bibitem{Klugel2015}
M.~Klugel and W.~Kellerer, ``Determining frequency reuse feasibility in
  device-to-device cellular networks,'' in \emph{IEEE Intl. Symposium on
  Personal, Indoor, and Mobile Radio Communications (PIMRC)}, Hong Kong,
  Aug.-Sep. 2015.

\bibitem{Xu2012}
C.~Xu, L.~Song, Z.~Han, Q.~Zhao, X.~Wang, and B.~Jiao, ``Interference-aware
  resource allocation for device-to-device communications as an underlay using
  sequential second price auction,'' in \emph{IEEE International Conference on
  Communications (ICC)}, Jun. 2012.

\bibitem{MWIS}
S.~Basagni, ``Finding a maximal weighted independent set in wireless
  networks,'' \emph{Telecommunication Systems}, vol.~18, no. 1-3, pp. 155--168,
  Sep. 2001.

\end{thebibliography}
\bibliographystyle{IEEEtran}

\end{document}